\documentclass[aps,prl,twocolumn,showpacs,superscriptaddress,groupedaddress]{revtex4-1}

\usepackage{graphicx}  
\usepackage{amsmath}
\usepackage{bm}        
\usepackage{amssymb}   

\usepackage{float}

\usepackage{color}

\begin{document}
\title{Slowing down supercooled liquids by manipulating their local structure}

\author{Susana Mar\'in Aguilar}
\affiliation{Laboratoire de Physique des Solides, CNRS, Universit\'e Paris-Sud, Universit\'e Paris-Saclay, 91405 Orsay, France}
\author{Henricus H. Wensink}
\affiliation{Laboratoire de Physique des Solides, CNRS, Universit\'e Paris-Sud, Universit\'e Paris-Saclay, 91405 Orsay, France}
\author{Giuseppe Foffi}
\email{giuseppe.foffi@u-psud.fr}
\affiliation{Laboratoire de Physique des Solides, CNRS, Universit\'e Paris-Sud, Universit\'e Paris-Saclay, 91405 Orsay, France}
\author{Frank Smallenburg}
\email{frank.smallenburg@u-psud.fr}
\affiliation{Laboratoire de Physique des Solides, CNRS, Universit\'e Paris-Sud, Universit\'e Paris-Saclay, 91405 Orsay, France}

\date{\today}
\begin{abstract}
Glasses remain an elusive and poorly understood state of matter. For example, it is not clear how we can control the macroscopic dynamics of  glassy systems by tuning the properties of their microscopic building blocks. In this paper, we propose a simple directional colloidal model that reinforces the optimal icosahedral local structure of binary hard-sphere glasses. We show that only this specific symmetry results in a dramatic slowing down of the dynamics.  Our results open the door to controlling the dynamics of dense glassy systems by selectively promoting specific local structural environments.
\end{abstract}
\maketitle

When we cool down or compress a liquid sufficiently rapidly to avoid crystallization, we end up with a glass: a dynamically arrested state which lacks long-range order. Glasses are ubiquitous in condensed matter and can be found in fields of great technological interest, including materials science, biomaterials, food science, and polymer physics. 
Although the transition from a fluid into a glass is dramatic from a dynamics perspective, it is accompanied by surprisingly little structural change: based on purely structural information, it is hard to tell whether a system has reached dynamical arrest or not \cite{Berthier2011,hunter2012physics}. The non-ergodic nature of glasses breaks the fundamental assumptions of equilibrium statistical mechanics and, as a consequence, it is extremely difficult to understand the microscopic origin of the glass transition.  For this reason, the bottom-up design of a glass former is not a trivial task.  In order to reliably reach a glassy state via cooling, the dynamics of the system have to slow down rapidly with decreasing temperature, such that the system avoids crystallization. Hence, the question is: how do we design particles with interactions capable of inducing slow dynamics? It is well understood that glassy dynamics are accompanied by the emergence of long-lived locally favored structures (LFS) \cite{tarjus2005frustration, royall2018local, malins2013identification, royall2015role, royall2017structures, shintani2006frustration, tanaka2003roles, miracle2007structural, doye2003favored}. However, these are always found a posteriori, by investigating models that are known to be good glass formers. In order to better understand  the microscopic mechanisms behind glass formation, it would be extremely useful to unravel how manipulation of the local structure of a glass former can be used to control its propensity for facilitating dynamical arrest.

In this Letter, we explore the dynamics of supercooled liquids of particles decorated with attractive patches on their surface, interpolating between the extreme cases of pure hard spheres and short-ranged attractive particles. Both these models have been widely studied experimentally, numerically and theoretically for their high-density glassy behavior \cite{Weeks2000, Berthier2005, berthier2016equilibrium, pham2002multiple,foffi2002evidence}. In particular, short-ranged attractive potentials display a remarkable reentrant glassy behavior \cite{bergenholtz1999nonergodicity, Dwason2001,Puertas2001, foffi2002evidence,Zaccarelli, pham2002multiple,eckert2002re,pham2004glasses}, in which the supercooled liquid first speeds up and then slows down again upon cooling.  Our directional model is inspired by the growing experimental availability of patchy colloidal particles  \cite{choueiri2016surface, yi2013recent,meester2016colloidal, biffi2015equilibrium, smalyukh2018liquid}. The ability of these particles to form well-defined directional bonds allows them to self-assemble into open crystal structures \cite{liu2016diamond, chen2011directed} and (at low densities) into strong network-forming glasses  \cite{biffi2015equilibrium, smallenburg2013liquids, de2006dynamics}. More importantly, carefully designed anisotropic interactions provide an effective route for manipulating the local structures of fluids, and have proven to be an effective way to suppress crystallization \cite{taffs2016role, molinero2006tuning, di2000off}.
Therefore, as one of the most versatile families of two-body anisotropic interactions, patchy particles are an ideal tool for pinning down the role of LFS in the vitrification of dense glasses.

Here, we use extensive computer simulations to demonstrate that directional interactions can be designed which selectively promote the LFS responsible for dynamical arrest. We find that that the glassy dynamics of patchy systems are remarkably similar for most patch geometries, but exhibit a dramatic slowdown for a select number of patch geometries which enhance icosahedral order, with 12-patch particles as the most extreme example. We demonstrate that these exceptional patch geometries are capable of matching and reinforcing the locally favored icosahedral structure of the fluid, providing an ideal route towards the design of systems which facilitate kinetic arrest. As icosahedral geometries also possess five-fold symmetry, known to suppress crystallization \cite{bernal1959geometrical, frank1952supercooling, karayiannis2011fivefold, taffs2016role}, patchy particles incorporating icosahedral symmetry are likely to be excellent glass formers.

\begin{figure}

\includegraphics[width=0.95\linewidth]{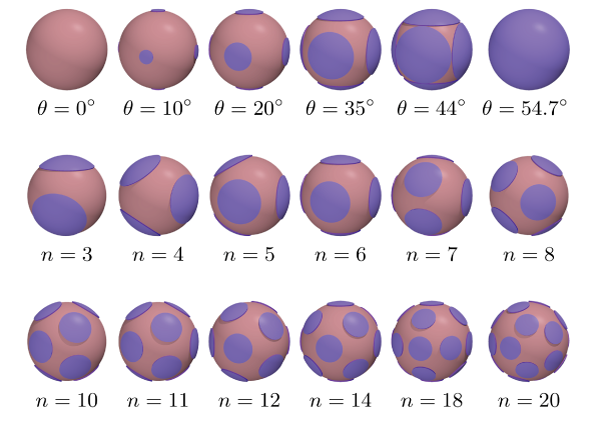} 

\caption{Illustration of the simulation model. The top row shows particles with 6 attractive patches with varying patch sizes determined by the opening angle $\theta$. The bottom row shows the patch geometry for selected numbers of patches. The patch placements correspond to the positions that maximize the minimum distance between two patches. Note that the 10, 11, and 12-patch geometries correspond to icosahedra with two, one, and zero vertices missing. 
}
\label{fig:model}
\end{figure}

To model the patchy interactions, we follow the Kern-Frenkel model \cite{Kern} and model the particles as hard spheres, decorated with $n$ attractive patches each. In order to suppress crystallization, we simulate a binary mixture of Kern-Frenkel particles of two sizes, with the size ratio given by $\sigma_S / \sigma_L = 0.833$, where $\sigma_{L(S)}$ denotes the size of the large (small) spheres. Two attractive patches form a bond with bonding energy $\epsilon$ when they overlap. The size of each patch is controlled by an opening angle $\theta$ and a fixed maximum interaction range $r_c = 1.031 \sigma_{ij}$, where $\sigma_{ij}$ is the contact distance between particles $i$ and $j$. This interaction range was chosen to be consistent with the square-well interaction in Ref. \onlinecite{Zaccarelli}. Here, we study systems with 3 to 20 patches, with the patches uniformly distributed over the surface. Specifically, we focus on to geometries where the minimum distance between two patches on the surface is maximized (sometimes referred to as spherical codes) \cite{sloane2000spherical}, as illustrated in Fig. \ref{fig:model}. Note that for 10, 11, and 12 patches, this results in patches placed on the vertices of an icosahedron, where for 10 and 11 patches there are two and one vertex omitted, respectively. We restrict our model to only allow a single bond for any pair of particles, even if multiple patches on the same particle overlap. With this restriction, the model interpolates between the hard-sphere model at $\theta = 0$ and the square-well model when $\theta$ is larger than a critical $\theta_f$ where the patches cover the entire sphere surface. We define the patch coverage fraction $\chi$ as the fraction of the surface of a particle which is covered by a patch. As long as the patches do not overlap, $\chi = n (1 - \cos \theta)/2$, with $n$ the number of patches. To simulate these particles, we use event-driven molecular dynamics (EDMD) simulations \cite{smallenburg2013liquids,hernandez2007discontinuous}. In each system, we fix the number of particles $N = 700$, the composition $x = 0.5$, and the packing fraction $\eta = 0.58$. The systems were equilibrated at fixed temperature for at least $10^4 \tau$, where $\tau = \sqrt{m \sigma_L^2/k_B T}$ is our time unit, $T$ is the temperature,  $m$ is the mass of a particle and $k_B$ is Boltzmann's constant. Followed by a production run at fixed energy for at least  $8 \cdot 10^4$ time units. In Ref. \onlinecite{Zaccarelli}, it was shown that this square-well system at this packing fraction has a clear reentrance as a function of temperature. Note that we report all dynamical quantities for the large particles only. For the small particles, the behavior is qualitatively the same.

\begin{figure}
\includegraphics[width=0.95\linewidth]{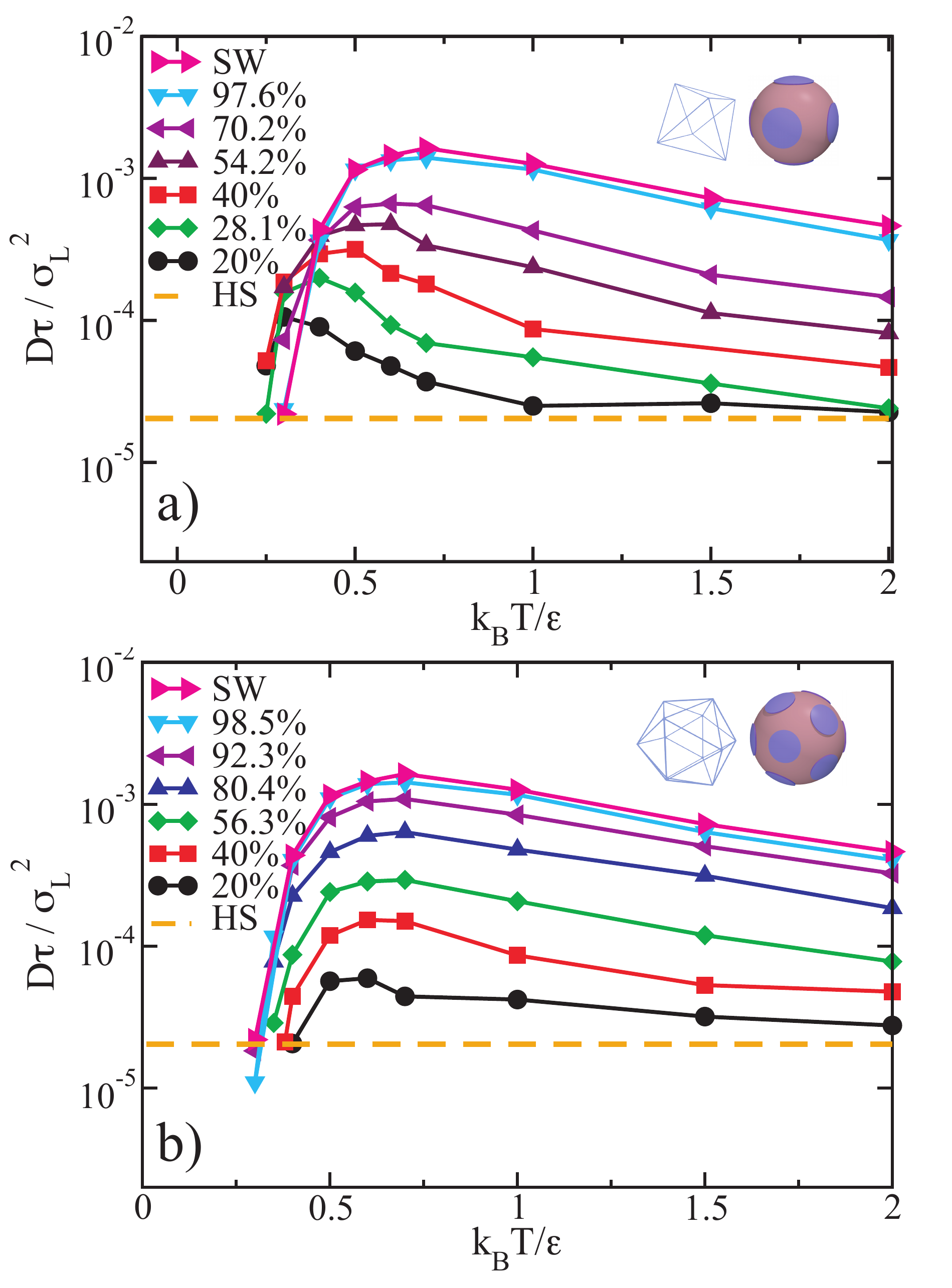} 
\caption{Translational diffusion coefficient calculated for a system of 6 patches (a) and 12 patches (b). Dashed lines correspond to the HS limit and the right triangles to the SW limit. The reentrance is conserved until $\chi$ is small enough and the system behaves like a mixture of HS. }
\label{fig:all_diff_coef}
\end{figure}

We begin our analysis by investigating whether the reentrant behavior of the diffusion coefficient as a function of temperature persists in the presence of directional interactions. To this end, we measure the dimensionless diffusion  coefficient $D \tau / \sigma_L^2$. In Fig. \ref{fig:all_diff_coef}a, we show this diffusion coefficient for systems with $n=6$ patches, as a function of the reduced temperature $k_B T / \epsilon$, for a number of different values of the patch coverage fraction $\chi$. In the limit of high $\chi$, we recover the case of an isotropic square-well model in the supercooled regime, and find reentrant diffusive behavior \cite{Sciortino, Zaccarelli}, where the system crosses over from an attractive glass to a repulsive one. Upon decreasing $\chi$, we observe an overall decrease in the diffusion, as shown in Fig. \ref{fig:all_diff_coef}a, which retains its reentrant behavior as observed in a recent mean-field solution for a simpler patch geometry \cite{yoshino2018translational}. However, the maximum in the diffusion rate shifts to lower temperatures as $\chi$ decreases.
This behavior can be understood from the observation that particles with lower coverage fractions form fewer bonds, implying that lower temperatures are required before bonding can similarly affect the dynamics (see SI). In the case of 12-patch particles (Fig. \ref{fig:all_diff_coef}b) we see a similar decrease in diffusivity by decreasing patch size, but the shift in the maximum is less pronounced. We observe similar trends for the other patch geometries we explored.

\begin{figure}

\includegraphics[width=0.95\linewidth]{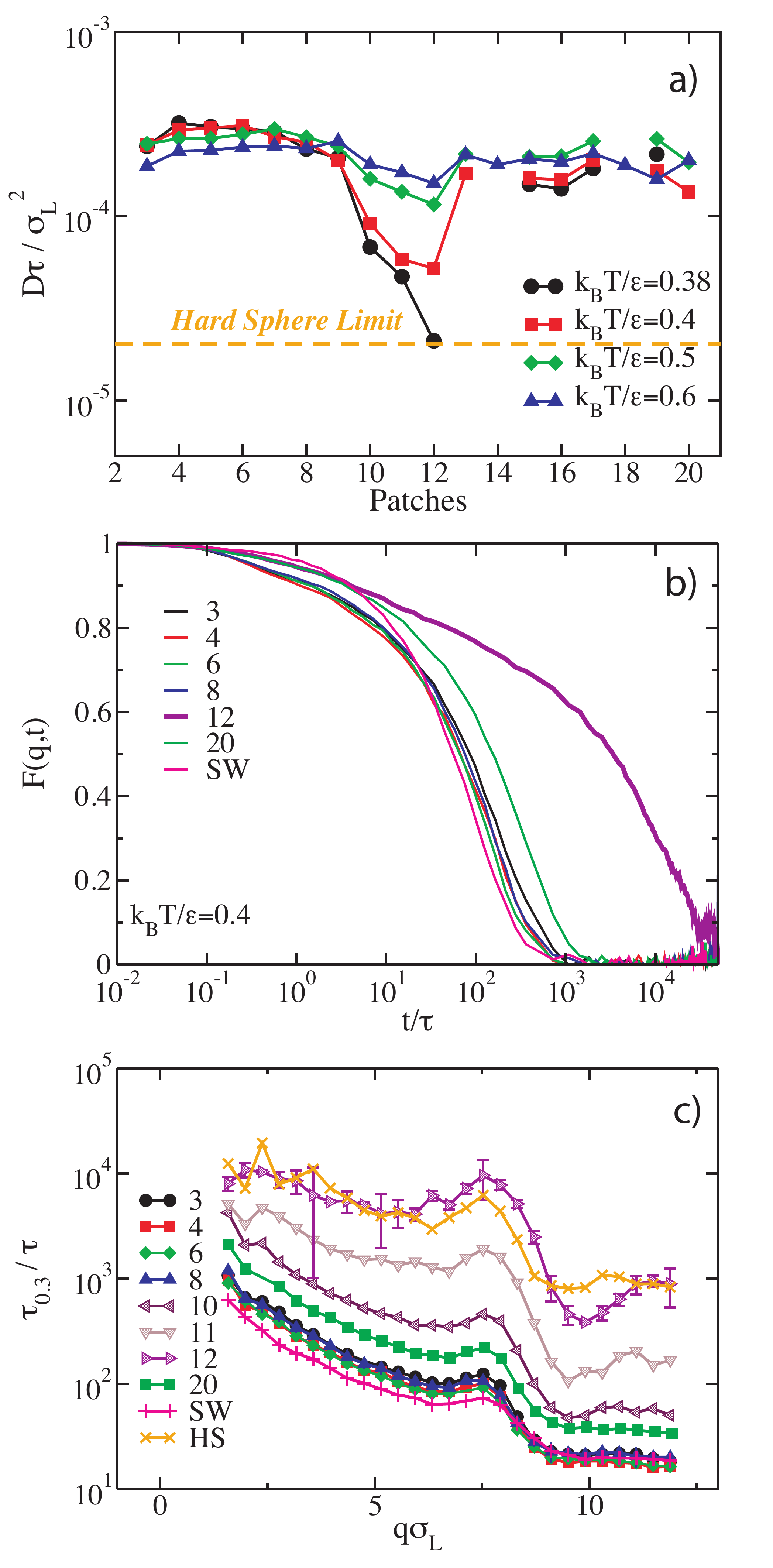} 

\caption{a) Diffusion coefficient as a function of number of patches, at different temperatures, for patchy particles with a fixed overall patch coverage fraction $\chi = 40\%$. Missing points correspond to crystallized systems.  b) Density correlation function of the same systems, for a wavevector $q$ corresponding to the first peak of the structure factor, at fixed temperature $k_B T / \epsilon = 0.4$. c) Wave vector dependence of the relaxation time $\tau_{0.3}$, at the same temperature. }
\label{fig:density_correlation}
\end{figure}

In order to explore the effect of the patch geometry in more detail, we now fix the total fraction of the particle surface covered by the patches to $\chi = 40\%$, and measure the diffusion constant as a function of the patch geometry. The results are shown in Fig. \ref{fig:density_correlation}a. Surprisingly, we see that the diffusion coefficient is largely independent of the patch geometry, except when the patch geometry matches icosahedral order. In the latter case, we instead see an extreme slowdown of the system at low temperature. Note that for several patch geometries (13, 14, 18, and 20-patch), we observe crystallization at the lowest temperatures investigated. These points have been omitted from Fig. \ref{fig:density_correlation}a. The observed crystal is always a close-packed crystal (i.e. a mixture of face-centered cubic and hexagonally close-packed) of the large spheres, except in the 20-patch case, where we find a CsCl crystal structure (see SI). For the other geometries, we never observe any signs of crystallization.

To make a more complete analysis of the dynamics, we measure in each system the density correlation function $F(q,t)$, which characterizes the relaxation time in the system at different length scales, given the wave vector $q$. In Fig. \ref{fig:density_correlation}b, we plot these correlation functions for temperature $k_B T / \epsilon = 0.4$, at the value of $q$ corresponding to the first peak in the structure factor. We observe an approximate collapse of the correlation functions for most geometries, while the 10, 11, and 12-patch systems are clear outliers which relax much more slowly. We define the relaxation time $\tau_{0.3}$ as the time at which the correlation function decays to 0.3. To confirm that the collapse of correlation functions is independent of $q$, we plot in Fig. \ref{fig:density_correlation}c the relaxation times as a function of $q$. Clearly, most systems relax at approximately the same rate, regardless of the length scale. However, the relaxation dynamics of the 10, 11, and 12-patch particles are much slower, with the 12-patch system being an order of magnitude slower.

To understand the origins of the dynamical slowdown in our system, we now focus on the local structure of the fluid. To this end, we analyze the prevalence of different local structural motifs in our systems using the Topological Cluster Classification (TCC) approach \cite{malins2013tcc}. This method detects local clusters of particles in particle configurations, based on the connectivity of the Voronoi construction. In particular, we look for TCC motifs matching either icosahedral order (13A) or crystalline order (FCC, HCP, or the 9X cluster matching body-centered cubic (BCC)). Note that the CsCl structure observed in 20-patch system is on a BCC lattice. Additionally, we identify icosahedral clusters in which the large and small particles are arranged such that they match the icosahedral motifs that occur in the Laves phases, which are known to be stable in binary hard sphere mixtures near this size ratio \cite{hynninen2009stability} (see SI). 

\begin{figure}
\includegraphics[width=0.95\linewidth]{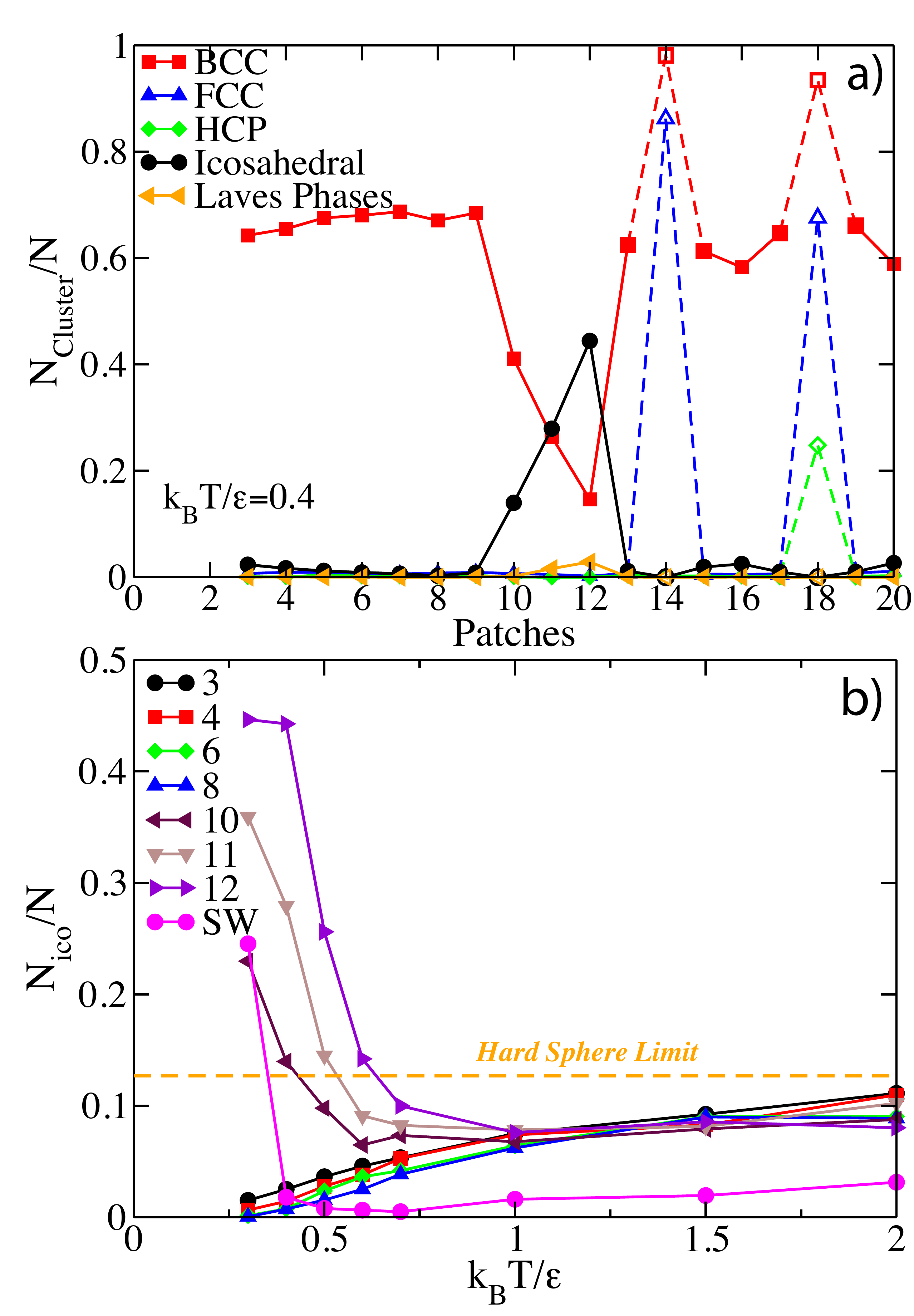}

\caption{Fraction of particles that are part of local structural motifs with specific symmetries, as obtained using Topological Cluster Classification \cite{malins2013tcc}. {\bf a)} Different structural motifs at a fixed temperature $k_B T /\epsilon = 0.4$. Note that the systems with 14 and 18 patches have partially crystallized into an FCC structure. {\bf b)} Particles in icosahedral clusters as a function of temperature for different patch numbers. The dashed horizontal line indicates the value in the hard-sphere limit.}
\label{fig:ico_measurement}
\end{figure}

In Fig. \ref{fig:ico_measurement}a, we show the fraction of particles that are part of at least one cluster of each type, for a fixed temperature $k_B T/\epsilon = 0.4$ and for all patch geometries. For the systems where crystallization occurred at this temperature (14-patch and 18-patch), we see large peaks in the populations of ``crystal-like'' clusters (FCC, BCC, HCP), as expected. In contrast, the other systems show little crystallinity, with the exception of the BCC motif, which is present in all fluids and does not indicate crystallinity on its own \cite{malins2013tcc}. Icosahedral order is only observed in significant amounts for the 10, 11, and 12-patch systems (which all match icosahedral symmetry), with the 12-patch system showing the largest fraction of icosahedral motifs. Intriguingly, these systems also show a strong drop in BCC ordering, and find only a negligible fraction of motifs corresponding to the Laves phases (i.e. a few isolated motifs of this type per configuration). 

In Fig. \ref{fig:ico_measurement}b, we plot the number of icosahedral motifs as a function of temperature for different patch geometries, as well as the square-well system. In the limit of high temperatures (i.e. the hard-sphere limit), we find that a significant fraction ($\approx 12\%$) of the particles are part of an icosahedral motif. As the temperature decreases, the number of icosahedral motifs initially decreases roughly identically for all patch geometries. In this regime, the patches essentially act, on average, as a weaker square well, and aid in the cage-breaking. However, at low temperatures, the 10, 11, and 12-patch systems start enhancing icosahedral order, reaching values well beyond the hard-sphere level, while all other patch geometries continue to further suppress icosahedral motifs. Interestingly, at very low temperatures, the square-well system also enhances icosahedral order, and does so quite suddenly, possibly hinting at a phase separation. 

Icosahedral local order has been linked to both dynamical slowdown \cite{jonsson1988icosahedral, cheng2008relationship, royall2015strong}, and to the suppression of crystallization \cite{bernal1959geometrical, frank1952supercooling, karayiannis2011fivefold,taffs2016role}, making it an ideal tool for the design of glassy materials. Here, we demonstrated that dynamical slowdown can be enhanced  by designing particle interactions which match and reinforce the long-lived local icosahedral structure responsible for caging in the repulsive glass \cite{leocmach2012roles,royall2015strong}. In contrast, for the 15 other patch geometries we explored, which do not match local icosahedral order, the attractions have surprisingly little impact on the dynamics of the system. Hence, our results suggest that -- for the spherical particles studied here -- icosahedral motifs are unique in their ability to slow down dynamics. This extreme sensitivity of dynamics to local icosahedral structures highlights the important role these motifs play in the kinetic arrest of disordered systems. Moreover, the inherent incompatibility of icosahedral order with long-ranged crystalline order suggests that these systems are also good glass formers from the perspective of avoiding crystallization. The main crystal structure found in our systems, an FCC lattice of large particles, is suppressed by five-fold symmetry \cite{taffs2016role} in hard-sphere systems. Indeed, simulations show that seeds of FCC (taken from a crystallized 18-patch system) melt back to a fluid in the 12-patch system (see SI).


Finally, it is interesting to note that at intermediate temperatures, all patchy interactions (as well as square-well interactions), lead to a disruption of the icosahedral cages which are present in the pure hard-sphere system. This suggests that the dynamical speedup that occurs at intermediate temperatures for all patch geometries can at least partly be interpreted as stemming from a disruption of the icosahedral cages of the repulsive glass. This is in contrast to lower temperatures, where the effect of interactions on dynamical slowdown hinges on their interplay with locally favored structures. Hence, by designing particles that inherently promote or disrupt the formation of ``slow'' local environments, we provide a clear recipe for controlling the dynamical slowdown of supercooled liquids.

We thank Francesco Sciortino for useful discussions. S. Mar\'in Aguilar acknowledges CONACyT for funding.

\bibliographystyle{apsrev4-1}
\bibliography{references_paper}
\end{document}